\batchmode
\makeatletter
\def\input@path{{"/home/jacob/Documents/Work/My Papers/2025-The ABL Rule and Weak Values/"}}
\makeatother
\documentclass[11pt,twoside,english]{article}
\usepackage[T1]{fontenc}
\usepackage[latin9]{inputenc}
\usepackage{babel}
\usepackage{amsmath}
\usepackage{amssymb}
\usepackage{geometry}
\usepackage{setspace}
\usepackage[pdftex,
 bookmarks=true,bookmarksnumbered=true,bookmarksopen=false,
 breaklinks=false,pdfborder={0 0 0},pdfborderstyle={},backref=false,colorlinks=false]
 {hyperref}
\hypersetup{
 pdfauthor={Jacob A. Barandes},
 pdfsubject={Physics},
 pdfkeywords={physics}}

\makeatletter

\let\originalleft\left
\let\originalright\right
\renewcommand{\left}{\mathopen{}\mathclose\bgroup\originalleft}
\renewcommand{\right}{\aftergroup\egroup\originalright}

\def\smalloverbrace#1{\mathop{\vbox{\m@th\ialign{##\crcr%
      \noalign{\kern3\p@}%
      \tiny\downbracefill\crcr\noalign{\kern3\p@\nointerlineskip}%
      $\hfil\displaystyle{#1}\hfil$\crcr}}}\limits}

\def\smallunderbrace#1{\mathop{\vtop{\m@th\ialign{##\crcr
   $\hfil\displaystyle{#1}\hfil$\crcr
   \noalign{\kern3\p@\nointerlineskip}%
   \tiny\upbracefill\crcr\noalign{\kern3\p@}}}}\limits}

\DeclareMathAlphabet{\mymathbb}{U}{bbold}{m}{n}

\makeatother

\begin{document}
\title{The ABL Rule and the Perils of Post-Selection}
\author{Jacob A. Barandes\thanks{Departments of Philosophy and Physics, Harvard University, Cambridge, MA 02138; jacob\_barandes@harvard.edu; ORCID: 0000-0002-3740-4418}
}
\date{\today}

\maketitle

\begin{abstract}
In 1964, Aharonov, Bergmann, and Lebowitz introduced their well-known
\textquoteleft ABL rule\textquoteright{} with the intention of providing
a time-symmetric formalism for computing novel kinds of conditional
probabilities in quantum theory. Later papers attached additional
significance to the ABL rule, including assertions that it supported
violations of the uncertainty principle. The present work challenges
these claims, as well as subsequent attempts to salvage the original
interpretation of the ABL rule. Taking a broader view, this paper
identifies a subtle category error at the heart of the ABL rule that
consists of confusing observables that belong to a single system with
emergent observables that arise only for physical ensembles. Along
the way, this paper points out other problems and fallacious reasoning
in the research literature surrounding the ABL rule, including the
misuse of post-selection, a reliance on pattern matching to classical
formulas, and a posture of \textquoteleft measurementism\textquoteright{}
that takes experimental data as providing answers to interpretational
questions.
\end{abstract}

\begin{center}
\global\long\def\quote#1{``#1"}%
\global\long\def\apostrophe{\textrm{'}}%
\global\long\def\slot{\phantom{x}}%
\global\long\def\eval#1{\left.#1\right\vert }%
\global\long\def\keyeq#1{\boxed{#1}}%
\global\long\def\importanteq#1{\boxed{\boxed{#1}}}%
\global\long\def\given{\vert}%
\global\long\def\mapping#1#2#3{#1:#2\to#3}%
\global\long\def\composition{\circ}%
\global\long\def\set#1{\left\{  #1\right\}  }%
\global\long\def\setindexed#1#2{\left\{  #1\right\}  _{#2}}%

\global\long\def\setbuild#1#2{\left\{  \left.\!#1\,\right|\,#2\right\}  }%
\global\long\def\complem{\mathrm{c}}%

\global\long\def\union{\cup}%
\global\long\def\intersection{\cap}%
\global\long\def\cartesianprod{\times}%
\global\long\def\disjointunion{\sqcup}%

\global\long\def\isomorphic{\cong}%

\global\long\def\setsize#1{\left|#1\right|}%
\global\long\def\defeq{\equiv}%
\global\long\def\conj{\ast}%
\global\long\def\overconj#1{\overline{#1}}%
\global\long\def\re{\mathrm{Re\,}}%
\global\long\def\im{\mathrm{Im\,}}%

\global\long\def\transp{\mathrm{T}}%
\global\long\def\tr{\mathrm{tr}}%
\global\long\def\adj{\dagger}%
\global\long\def\diag#1{\mathrm{diag}\left(#1\right)}%
\global\long\def\dotprod{\cdot}%
\global\long\def\crossprod{\times}%
\global\long\def\Probability#1{\mathrm{Prob}\left(#1\right)}%
\global\long\def\Amplitude#1{\mathrm{Amp}\left(#1\right)}%
\global\long\def\cov{\mathrm{cov}}%
\global\long\def\corr{\mathrm{corr}}%

\global\long\def\absval#1{\left\vert #1\right\vert }%
\global\long\def\expectval#1{\left\langle #1\right\rangle }%
\global\long\def\op#1{\hat{#1}}%

\global\long\def\bra#1{\left\langle #1\right|}%
\global\long\def\ket#1{\left|#1\right\rangle }%
\global\long\def\braket#1#2{\left\langle \left.\!#1\right|#2\right\rangle }%

\global\long\def\parens#1{(#1)}%
\global\long\def\bigparens#1{\big(#1\big)}%
\global\long\def\Bigparens#1{\Big(#1\Big)}%
\global\long\def\biggparens#1{\bigg(#1\bigg)}%
\global\long\def\Biggparens#1{\Bigg(#1\Bigg)}%
\global\long\def\bracks#1{[#1]}%
\global\long\def\bigbracks#1{\big[#1\big]}%
\global\long\def\Bigbracks#1{\Big[#1\Big]}%
\global\long\def\biggbracks#1{\bigg[#1\bigg]}%
\global\long\def\Biggbracks#1{\Bigg[#1\Bigg]}%
\global\long\def\curlies#1{\{#1\}}%
\global\long\def\bigcurlies#1{\big\{#1\big\}}%
\global\long\def\Bigcurlies#1{\Big\{#1\Big\}}%
\global\long\def\biggcurlies#1{\bigg\{#1\bigg\}}%
\global\long\def\Biggcurlies#1{\Bigg\{#1\Bigg\}}%
\global\long\def\verts#1{\vert#1\vert}%
\global\long\def\bigverts#1{\big\vert#1\big\vert}%
\global\long\def\Bigverts#1{\Big\vert#1\Big\vert}%
\global\long\def\biggverts#1{\bigg\vert#1\bigg\vert}%
\global\long\def\Biggverts#1{\Bigg\vert#1\Bigg\vert}%
\global\long\def\Verts#1{\Vert#1\Vert}%
\global\long\def\bigVerts#1{\big\Vert#1\big\Vert}%
\global\long\def\BigVerts#1{\Big\Vert#1\Big\Vert}%
\global\long\def\biggVerts#1{\bigg\Vert#1\bigg\Vert}%
\global\long\def\BiggVerts#1{\Bigg\Vert#1\Bigg\Vert}%
\global\long\def\ket#1{\vert#1\rangle}%
\global\long\def\bigket#1{\big\vert#1\big\rangle}%
\global\long\def\Bigket#1{\Big\vert#1\Big\rangle}%
\global\long\def\biggket#1{\bigg\vert#1\bigg\rangle}%
\global\long\def\Biggket#1{\Bigg\vert#1\Bigg\rangle}%
\global\long\def\bra#1{\langle#1\vert}%
\global\long\def\bigbra#1{\big\langle#1\big\vert}%
\global\long\def\Bigbra#1{\Big\langle#1\Big\vert}%
\global\long\def\biggbra#1{\bigg\langle#1\bigg\vert}%
\global\long\def\Biggbra#1{\Bigg\langle#1\Bigg\vert}%
\global\long\def\braket#1#2{\langle#1\vert#2\rangle}%
\global\long\def\bigbraket#1#2{\big\langle#1\big\vert#2\big\rangle}%
\global\long\def\Bigbraket#1#2{\Big\langle#1\Big\vert#2\Big\rangle}%
\global\long\def\biggbraket#1#2{\bigg\langle#1\bigg\vert#2\bigg\rangle}%
\global\long\def\Biggbraket#1#2{\Bigg\langle#1\Bigg\vert#2\Bigg\rangle}%
\global\long\def\angs#1{\langle#1\rangle}%
\global\long\def\bigangs#1{\big\langle#1\big\rangle}%
\global\long\def\Bigangs#1{\Big\langle#1\Big\rangle}%
\global\long\def\biggangs#1{\bigg\langle#1\bigg\rangle}%
\global\long\def\Biggangs#1{\Bigg\langle#1\Bigg\rangle}%

\global\long\def\vec#1{\mathbf{#1}}%
\global\long\def\vecgreek#1{\boldsymbol{#1}}%
\global\long\def\idmatrix{\mymathbb{1}}%
\global\long\def\projector{P}%
\global\long\def\permutationmatrix{\Sigma}%
\global\long\def\densitymatrix{\rho}%
\global\long\def\krausmatrix{K}%
\global\long\def\stochasticmatrix{\Gamma}%
\global\long\def\lindbladmatrix{L}%
\global\long\def\dynop{\Theta}%
\global\long\def\timeevop{U}%
\global\long\def\hadamardprod{\odot}%
\global\long\def\tensorprod{\otimes}%

\global\long\def\inprod#1#2{\left\langle #1,#2\right\rangle }%
\global\long\def\normket#1{\left\Vert #1\right\Vert }%
\global\long\def\hilbspace{\mathcal{H}}%
\global\long\def\samplespace{\Omega}%
\global\long\def\configspace{\mathcal{C}}%
\global\long\def\phasespace{\mathcal{P}}%
\global\long\def\spectrum{\sigma}%
\global\long\def\restrict#1#2{\left.#1\right\vert _{#2}}%
\global\long\def\from{\leftarrow}%
\global\long\def\statemap{\omega}%
\global\long\def\degangle#1{#1^{\circ}}%
\global\long\def\trivialvector{\tilde{v}}%
\global\long\def\eqsbrace#1{\left.#1\qquad\right\}  }%
\global\long\def\operator#1{\operatorname{#1}}%
\par\end{center}

\section{Introduction\label{sec:Introduction}}

In 1964, Aharonov, Bergmann, and Lebowitz (ABL) published a highly
influential paper titled ``Time Symmetry in the Quantum Process of
Measurement'' in the journal \emph{Physical Review} (Aharonov, Bergmann,
Lebowitz, 1964)\nocite{AharonovBergmannLebowitz:1964tsitqpom}. According
to the \emph{Physical Review} website, the ABL paper now has over
700 citations, which include papers from a variety of areas: quantum
foundations (Griffiths 1984)\nocite{Griffiths:1984chatioqm}, quantum
cosmology (Gell-Mann, Hartle 1994)\nocite{Gell-MannHartle:1994tsaaiqmaqc},
closed timelike curves (Lloyd et al. 2011)\nocite{LloydMacconeGarcia-PatronGiovannettiShikano:2011qmotttpst},
and black holes (Horowitz, Maldacena; Lloyd 2006; Harlow 2016; Akers
et al. 2024)\nocite{HorowitzMaldacena:2004tbhfs,Lloyd:2006acefbhifspm,Harlow:2016jlobhaqi,AkersEngelhardtHarlowPeningtonVardhan:2024tbhifnicac}.
It is worth noting that all these cited papers not only refer to the
ABL paper as a whole, but explicitly refer to its claims at having
provided a time-symmetric formulation of quantum theory.

Importantly, the ABL paper initiated the widespread use of post-selection
in quantum theory. It also inspired the development of weak values
in a 1988 paper by Aharonov, Albert, and Vaidman that has since received
over 2,200 citations (Aharonov, Albert, Vaidman 1988)\nocite{AharonovAlbertVaidman:1988htroamoacotsoas12pctotb1}.\footnote{A search using Google Scholar for \{\textquotedblleft quantum\textquotedblright{}
AND (``postselection'' OR ``post-selection\textquotedblright{}
OR ``post-select'' OR ``postselect'' OR ``post-selected'' OR
``postselected'')\} yields no valid results from before 1964. For
the span of years stretching from 1964 to 2025, it yields almost 19,000
results, the vast majority after the year 2000. A search for \{``quantum''
AND (``weak value'' OR ``weak values'')\} yields over 6,000 results.}

The main result of the ABL paper was its derivation of the \textquoteleft ABL
rule,\textquoteright{} a formula for a certain class of conditional
probabilities that purportedly gives a time-symmetric formulation
of quantum theory. The present work will clarify the precise meaning
of the conditional probabilities that the ABL paper introduced, and
argue that the ABL paper's formulation was not, in fact, time symmetric,
but symmetric under a conceptually different class of transformations:
chronological reverse-orderings of measurement sequences. By way of
analogy, consider the distinction between, on the one hand, turning
over a deck of face-down playing cards so that they are now all face-up,
and, on the other hand, separating out all the playing cards and then
re-stacking them in the opposite order while keeping them face-down
the entire time.

The present work will also critique later papers that have attempted
to salvage the original time-symmetric interpretation of the ABL rule,
or that have tried to extend the ABL rule in ways that lead to supposed
violations of the uncertainty principle.  Other work will argue that
weak values do not provide supporting evidence for the arguments made
in the ABL paper, and run into fundamental interpretational difficulties
of their own (Barandes 2026)\nocite{Barandes:2025ttwwv}.

More broadly, this paper will argue that post-selection, whether used
in the context of the ABL rule or for other purposes in quantum-physics
research, is not an innocent or innocuous procedure, but routinely
leads to statistical artifacts that are often mistakenly attributed
to the exotic nature of quantum theory. A corollary is a call for
research journals to insist that authors who rely on post-selection
to obtain surprising results should explain why those results are
features of quantum mechanics itself and not merely artifacts of post-selection.

With all that said, this paper will not argue that all research inspired
by the ABL paper is wrong or does not work. To the contrary, the
ABL paper has inspired  new theoretical and experimental frameworks
that stand on their own merits and do not depend on unsupported interpretational
claims.

Section~\ref{sec:Preliminaries} will cover various preliminary topics
that will be important for this paper's critical treatment of the
ABL rule, including relevant foundational concepts in textbook quantum
theory, fallacies related to ensembles and post-selection, and fallacies
related to pattern matching and \textquoteleft measurementism.\textquoteright{}
Section~\ref{sec:The-ABL-Rule} will present a careful treatment
of the ABL rule in detail, starting with a first look at the ABL rule
itself, followed by a discussion of the role played by boundary conditions,
a technical derivation of the ABL rule, an analysis of time symmetry
in the ABL rule, a rigorous analysis of the original ABL paper, an
assessment of historical attempts to justify key claims surrounding
the ABL rule, and a novel perspective on arguments that the ABL rule
provides a loophole in the uncertainty principle. Section~\ref{sec:Conclusion}
will conclude with a summary and a discussion of larger ramifications.

\section{Preliminaries\label{sec:Preliminaries}}

\subsection{Textbook Quantum Theory\label{subsec:Textbook-Quantum-Theory}}

A sufficiently precise deconstruction of the claims made by the ABL
paper and subsequent research will require engaging with the axiomatic
foundations of orthodox or textbook quantum theory. To be self-contained
and to put all the cards on the table, here is a retelling of the
standard axioms for the textbook theory, as laid down by Dirac (1930)
and von Neumann (1932)\nocite{Dirac:1930pofm,vonNeumann:1932mgdq}
(DvN):
\begin{enumerate}
\item Quantum states: A quantum system is represented at any moment in time
by a quantum state associated with a Hilbert space of some finite
or infinite dimension. In general, a quantum state is a unit-trace,
self-adjoint, positive semidefinite density matrix, or density operator,
$\densitymatrix$. If the density matrix is rank-one, then one can
instead use a unit vector defined up to arbitrary overall phase, called
a state vector, or wave function, $\ket{\Psi}$. Sometimes one adds
a mereological postulate that the Hilbert spaces of composite quantum
systems are the tensor products of the respective Hilbert spaces of
their constituent subsystems, where the quantum states of subsystems
are related to the quantum states of their composite systems by the
partial-trace operation.
\item Unitary time evolution: If a quantum system is a closed system, meaning
that it is isolated from mutual interactions with any other systems,
then the system's quantum state evolves according to a time-indexed
family of unitary operators, which collectively define a time-dependent
unitary operator known as the system's time-evolution operator. Under
appropriate smoothness assumptions, one can express unitary time evolution
as a differential equation that is first-order in time\textemdash the
von Neumann equation for density matrices or the Schrödinger equation
for state vectors.
\item Observables: A single quantum system has an associated set of observables.
Each observable is represented by a self-adjoint operator $A=A^{\adj}$
on the system's Hilbert space, where the possible numerical measurement
outcomes that make up the observable's spectrum correspond to the
eigenvalues of that self-adjoint operator. (More generally, one can
work with positive-operator-valued measures, or POVMs.)
\item The Born rule: To compute the probability with which a measurement
of an observable will yield an outcome belonging to some collection
of eigenvalues of the associated self-adjoint operator, one projects
the quantum state down to the subspace corresponding to that set of
eigenvalues, and then one computes the trace of the resulting operator,
or the norm-square of the resulting vector.
\item Collapse: Immediately following this measurement outcome, one projects
or collapses the system's quantum state down to the appropriate subspace
and then renormalizes it so that it has trace or norm equal to $1$
again.
\end{enumerate}

One should not take this paper's use of the DvN axioms as an endorsement.
The DvN axioms famously do not define what precisely counts as a measurement,
leading to ambiguities over when to apply the second axiom (unitary
time evolution) and when to apply the fifth axiom (collapse). Unitary
time evolution alone is not able to single out unique measurement
outcomes, even with the invocation of open-system dynamics and decoherence.
These difficulties make up the famous measurement problem. That said,
most other axiomatic formulations and interpretations of quantum theory
make essentially equivalent predictions for the kinds of experimental
protocols that will be examined in this paper, so the DvN axioms will
suffice.

\subsection{Ensembles and Post-Selection\label{subsec:Ensembles-and-Post-Selection}}

A central theme in this paper's critical analysis will be the distinction
between single-system observables, which are empirically accessible
features of a single system, and ensemble observables, which are irreducibly
emergent features of a physical ensemble as a whole. As emphasized,
for instance, by Hance, Rarity, and Ladyman (2023, Section VI)\nocite{HanceRarityLadyman:2023wvatpoaqp},
ensemble observables are categorically different from single-system
observables\textemdash even if, in practice, it may sometimes be necessary
to employ an ensemble to \emph{study} the observables of a single
system. That is, the mere use of an ensemble to probe a single-system
observable does not make that observable an ensemble observable, because
ensemble observables are a categorically distinct notion. For example,
in quantum theory, the eigenvalue spectrum of an observable is a single-system
concept, even if an experimentalist might use a physical ensemble
to survey that spectrum fully.

A corollary is that it is a category error to try to identify single-system
observables with ensemble observables, or to draw direct inferences
about single-system observables from ensemble observables. This category
error will be called the ensemble fallacy: 

\begin{equation}
\left.\begin{minipage}{\columnwidth}
\leftskip=10pt 
\rightskip=60pt 

The Ensemble Fallacy: The category error of attempting to identify
a single-system observable with an ensemble observable, or to draw
direct inferences about a single-system observable from an ensemble
observable without independent rigorous justification.\label{eq:DefEnsembleFallacy}

\end{minipage}\hspace{-50pt}\right\}
\end{equation}

It is easy to confuse an emergent ensemble observable with a single-system
observable, again because ensembles are sometimes used in practice
to gain empirical access to single-system observables. A simple example
might therefore be worthwhile here. 

A star has various observable features, like mass, volume, electromagnetic
spectrum, chemical composition, age, location, and so forth. One can
certainly study these single-star observables by using ensembles\textemdash say,
by looking at lots of stars of similar mass at different moments during
their life cycle. For millennia, it has also been the case that human
societies have selected ensembles of stars in the sky that form various
mythologically inspired shapes that we know as constellations. A constellation
is an ensemble observable that depends on arbitrary choices made by
people, and is not a single-star observable. The stars that make up
a constellation are typically far apart in three-dimensional space,
and have essentially no mutual interactions. The fact that the constellation
Orion refers to a hunter tells us exceedingly little about the observables
of any of its individual stars. It certainly does not suggest that
each star contains a small degree of \textquoteleft hunter-ness\textquoteright{}
among its observables.\footnote{This particular example does not feature a cooperative form of emergence,
because the stars essentially do not interact with each other. By
contrast, in the case of superradiance, an ensemble of systems interact
with each other to produce an irreducibly emergent pattern of radiation.
However, superradiance, like a constellation, is not a single-system
observable, and to confuse it with one would be to commit the ensemble
fallacy.}

One can also define ensembles by post-selection, which is a key component
of the ABL rule. Post-selection refers to culling\textemdash or, perhaps
less politely, cherry-picking\textemdash members of an ensemble after
the fact, meaning after the experimental procedure, with the effect
of changing the ensemble's statistical properties. Post-selection
is a delicate issue in statistics, and can easily lead one to draw
erroneous conclusions about a set of data, an effect that is a form
of selection bias (Hernán, Hernández-\'{D}\i az, Robins 2004)\nocite{HernanHernandez-DiazRobins:2004asatsb}.
Indeed, one goal of many statisticians is to \emph{avoid }or \emph{correct}
for post-selection, as a way to escape or mitigate its potential
consequences, whereas for the ABL rule, one is supposed to \emph{implement}
post-selection quite deliberately. Also, by construction, post-selection
produces ensemble observables, not single-system observables, because
the post-selection criteria are entirely up to whoever does the culling
or cherry-picking and are inherently statements about the ensemble
as a whole. It will be helpful to give this general class of errors
a name:

\begin{equation}
\left.\begin{minipage}{\columnwidth}
\leftskip=10pt 
\rightskip=60pt 

The Post-Selection Fallacy: Drawing erroneous conclusions about a
system or ensemble due to the implicit or explicit use of post-selection.\label{eq:DefPostSelectionFallacy}

\end{minipage}\hspace{-50pt}\right\}
\end{equation}

One can easily come up with examples that illustrate both the ensemble
fallacy and the post-selection fallacy:
\begin{itemize}
\item Consider a classical ensemble of experimental trials consisting
of identical coins with the unusual observable feature that each time
a coin lands on tails, it becomes 1\% darker in appearance. In each
trial, a coin is tossed 100 times in a row, for the purpose of estimating
whether the coins are fair. If the ensemble of trials is post-selected
on the coins being, say, at least 70\% dark at the end, then the post-selected
ensemble will under-sample coins that land frequently on heads. Any
skewing of the final frequency distribution due to this post-selection
decision is obviously not an intrinsic feature of the individual coins,
but an ensemble-level statistical artifact of the arbitrary choice
of post-selection. In particular, if the final frequency distribution
of heads and tails fails to be 50/50, then one would be incorrect
to conclude that the coins were not fair.
\item Consider a health symptom $S$ that arises if and only if a patient
has a condition $A$, a condition $B$, or both. Suppose that the
conditions $A$ and $B$ are statistically independent in the general
population, and that each condition occurs with a probability of 10\%.
In terms of the probabilities $p_{A}$ for $A$, $p_{B}$ for $B$,
and $p_{AB}$ for the logical conjunction of $A$ \emph{and} $B$,
one therefore has 
\[
p_{A}=\frac{1}{10},\quad p_{B}=\frac{1}{10},\quad p_{AB}=p_{A}p_{B}=\frac{1}{100}.
\]
 It follows that if one imagines an ensemble of, say, $N=10,\!000$
people in the general population, then one would expect that $N_{A}\approx p_{A}N=1,\!000$
have the condition $A$, that $N_{B}\approx p_{B}N=1,\!000$ have
the condition $B$, and that $N_{AB}\approx p_{AB}N=100$ have both
conditions. However, suppose that one post-selects on patients who
have the health symptom $S$. By assumption, the number of such patients
is $N_{S}=N_{A}+N_{B}-N_{AB}\approx1,\!900$, where the subtraction
of $N_{AB}$ avoids double-counting patients who have both the conditions
$A$ and $B$. From among this post-selected ensemble, one sees that
$N_{A}/N_{S}\approx10/19$ have $A$, that $N_{B}/N_{S}\approx10/19$
have $B$, and that $N_{AB}/N_{S}\approx1/19$ have both $A$ and
$B$. However, $1/19\ne10/19\times10/19$, so the post-selected sample
shows a (negative) statistical correlation between $A$ and $B$.
This correlation, however, is an ensemble-level statistical artifact
of the choice of post-selection, and does not reflect a true interdependence
between $A$ and $B$. This example is an instance of Berkson's paradox
(Berkson 1946)\nocite{Berkson:1946lotaoftathd}, or collider bias
(Cole et al., 2009)\nocite{ColePlattSchistermanChuWestreichRichardsonPoole:2009ibdtcoac}.
\item Consider an ensemble of trials involving stones and holes. In each
trial, one tosses a small stone in a slightly random direction toward
a collection of holes. Suppose, further, that in each trial, precisely
one hole, picked at random, is completely covered up by a metal cover
or shutter. If the ensemble is post-selected to include only trials
in which a metallic clanging noise is recorded, then one will find
that every stone in the post-selected ensemble of trials has been
blocked from entering a hole. It would obviously be incorrect to conclude
from this statistical analysis that the shutter itself possesses the
mysterious capacity to cover all the holes at once. The hole-blocking
is an emergent property of the post-selected ensemble, and not a single-system
property of the shutter, so making inferences about the shutter from
the post-selected ensemble would be to commit the ensemble fallacy
\eqref{eq:DefEnsembleFallacy}. The choice of this specific example
was not accidental\textemdash it is closely related to the subject
of several papers on the use of the ABL rule for quantum systems (Aharanov,
Vaidman 2003; Kastner 2004)\nocite{AharonovVaidman:2003hosccns,Kastner:2004sbbnptspiqt}.
\item Consider an ensemble of trials in which an object can be placed in
any of three boxes. In each trial, there is only one object. Suppose,
moreover, that with 50/50 odds, a detection machine will check either
the first box for the object, or the second box, but not both. The
machine will show a bright green light if it succeeds in finding the
object, and the machine never checks the third box in any of the trials.
If the trials are post-selected on the logically conjunctive proposition
that the machine has checked the first box and ends up displaying
its green light, then, among the members of the resulting ensemble,
the odds of the object being in the first box are 100\%. If the trials
are instead post-selected on the logical conjunction of the machine
checking the second box and showing its green light, then, among the
members of \emph{that} ensemble, the odds of the object being in the
second box are likewise 100\%. However, despite finding 100\% odds
in both cases, one would be incorrect to conclude from this experiment
that the object was somehow in \emph{both} the first and second boxes.
Again, this example was chosen for a reason\textemdash a very similar
example became a significant focus of interest in the research literature
on the ABL rule (Aharonov, Albert, D'Amato 1985; Aharonov, Vaidman
1991; Cohen 1995; Vaidman 1996a; Vaidman 1998; Kastner 1999)\nocite{AlbertAharonovDAmato:1985cnspoqm,AharonovVaidman:1991cdoaqsaagt,Cohen:1995papqscmach,Vaidman:1996wmeor,Vaidman:1998tsqt,Kastner:1999ttbpaortrtcuotar}.
\end{itemize}

This paper will show that the ABL rule similarly refers to emergent
observables of physical ensembles that cannot be identified as single-system
observables and cannot be used to draw direct inferences about single-system
observables. To make such identifications or inferences would, again,
be precisely to commit the ensemble fallacy \eqref{eq:DefEnsembleFallacy}.
This paper will also argue that some claimed implications of the ABL
rule run aground on the post-selection fallacy \eqref{eq:DefPostSelectionFallacy}.
Other work will argue that weak values suffer from similar problems
(Barandes 2026)\nocite{Barandes:2025ttwwv}. 

Again, none of this criticism should be taken to mean that all the
spin-offs of the ABL rule should be discarded. Ensemble observables
are still observables, after all, and have their uses. Indeed, constellations
have helped people navigate the world for many generations (Huth 2013)\nocite{Huth:2013tlaofow}.
Those successful applications of constellations, however, have not
revealed very much about the intrinsic nature of individual stars.

\subsection{Pattern Matching and Measurementism\label{subsec:Pattern-Matching-and-Measurementism}}

Beyond the ensemble and post-selection fallacies introduced in Subsection~\ref{subsec:Ensembles-and-Post-Selection},
the present work will identify two other, larger problems with the
ABL paper and related research literature. The first is an overuse
of pattern matching to classical formulas or interpretations in quantum-physics
research. The second is \textquoteleft measurementism,\textquoteright{}
which will refer to the philosophical posture that if an experiment
ends up obtaining a specific quantity, then that experiment alone
confirms a previously favored interpretation of that quantity. (``We
have measured it many times in the lab, so how could it be wrong?''
or ``The fact that our theory leads to an experiment that can be
performed is enough to justify the theory.'')

In this paper, pattern matching will refer to the common practice
of trying to impose, by decree, relationships between classical notions
and quantum notions. This practice is especially common when the relevant
notions play similar or analogous \emph{functional} roles in classical
physics and in quantum physics, despite consisting of fundamentally
different \emph{mathematical} structures.

One specific form of pattern matching consists of looking at classical
formulas that have reasonably transparent conceptual or physical meanings,
and then trying to guess corresponding formulas in the quantum case
by formal analogy\textemdash say, by replacing random variables with
self-adjoint operators, replacing Poisson brackets with commutators,
replacing classical probability distributions with density matrices,
replacing marginalization with partial traces, or replacing stochastic
processes with quantum channels. Dirac's canonical quantization is
explicitly a form of pattern matching of this kind, and a great deal
of ongoing research in quantum causal modeling involves explicit pattern
matching to the ingredients of classical causal models.

Notice that in each pair of mathematical objects in the previous paragraph\textemdash for
instance, random variables and self-adjoint operators\textemdash the
functional role played by the latter member in quantum physics is
similar or analogous to the functional role played by the former member
in classical physics. However, the two members of each pair are based
on fundamentally different mathematical structures.\footnote{By contrast, if two physical theories each contain a feature that,
while perhaps playing different functional roles in the two theories,
turn out to consist of sufficiently similar mathematical structures
in both theories, then one may be justified in learning about the
feature of the first theory from the corresponding feature of the
other theory. Importantly, this case arises in the use of \textquoteleft analogue
models,\textquoteright{} for which one indirectly studies a feature
of an empirically inaccessible lower-level theory, like quantum theory
or general relativity, by examining or running measurements that probe
mathematically similar features of an empirically accessible higher-level
theory.}

A different form of pattern matching consists of taking a formula
already obtained in the quantum case and then assigning it an interpretation
or meaning through an appeal to a similar-looking classical formula.
An example with particular relevance to the ABL paper would be assuming
that the initial and final boundary conditions used in Lagrangian
mechanics are analogous to pre-selections and post-selections in quantum
theory, as will be discussed in Subsection~\ref{subsec:Boundary-Conditions}.

These forms of pattern matching, however, are never justified on
their own. While pattern matching can be a useful heuristic or provisional
first step, one must ultimately be able to derive any claimed formula
in quantum theory either from the DvN axioms, as reviewed in Subsection~\ref{subsec:Textbook-Quantum-Theory},
or from some rigorously self-consistent modification or replacement
of the DvN axioms. Anything else would lie strictly outside of an
axiomatic framework, and so would either need to be based on a direct
appeal to scientific induction or abduction, or would essentially
require engaging in a form of hand-waving.

It will be convenient to give these illicit forms of pattern matching
a name:

\begin{equation}
\left.\begin{minipage}{\columnwidth}
\leftskip=10pt 
\rightskip=60pt 

The Pattern-Matching Fallacy: Declaring the validity of a new mathematical
construct in quantum theory by analogy with a known classical mathematical
construct, or assigning an interpretation to an existing mathematical
construct in quantum theory by analogy with the interpretation of
a known classical mathematical construct.\label{eq:DefPatternMatchingFallacy}

\end{minipage}\hspace{-50pt}\right\}
\end{equation}

When even pattern matching is insufficient to vindicate or justify
a claimed interpretation of some physical quantity, another tempting
option might be \textquoteleft measurementism,\textquoteright{} a
pervasive philosophical attitude that this paper will define as the
following fallacy:

\begin{equation}
\left.\begin{minipage}{\columnwidth}
\leftskip=10pt 
\rightskip=60pt 

The Measurementist Fallacy (or Measurementism): If a quantity of ambiguous
interpretation can be measured experimentally, then experiments alone
can provide confirmatory support for a favored interpretation of that
quantity, or a justification for theoretical work that led to the
consideration of that quantity.\label{eq:DefMeasurementistFallacy}

\end{minipage}\hspace{-50pt}\right\}
\end{equation}

\noindent As an immediate application, the mere fact that a specific
frequency ratio appearing in the ABL rule is experimentally measurable\textemdash and,
indeed, has been measured in the laboratory in specific cases\textemdash does
not vindicate many of the strong interpretational claims that have
been made about that frequency ratio or about the ABL rule.

\section{The ABL Rule\label{sec:The-ABL-Rule}}

\subsection{A First Look at the ABL Rule\label{subsec:A-First-Look-at-the-ABL-Rule}}

At its core, the ABL paper was concerned with the application of quantum
theory to a specific kind of experimental protocol.

Consider a large ensemble of $N\gg1$ identical quantum systems with
negligible internal time evolution. Suppose that for each system in
the ensemble, an external agent or observer measures $n+2$ possibly
distinct observables $A,C_{1},\dots,C_{n},B$ in succession, while
sequentially recording the associated measurement outcomes. For the
purposes of this experimental protocol, assume that each observable
is complete, in the sense that its measurement outcome, together with
the DvN collapse axiom reviewed in Subsection~\ref{subsec:Textbook-Quantum-Theory},
yields a single, specific state vector. The observer counts up the
number $N_{ab}$ of members of the ensemble that share a specific
pre-selected measurement outcome $a$ for the first observable $A$
and a specific post-selected measurement outcome $b$ for the last
observable $B$. Then, from that subensemble labeled by the pair $a$
and $b$, the observer counts up the number $N_{ac_{1}\cdots c_{n}b}\leq N_{ab}$
of members that share a specific sequence $c_{1},\dots,c_{n}$ of
measurement results for the $n$ respective intermediate observables
$C_{1},\dots,C_{n}$. The numerical fraction 
\begin{equation}
0\leq\frac{N_{ac_{1}\cdots c_{n}b}}{N_{ab}}\leq1\label{eq:ABLFraction}
\end{equation}
 defines a specific, experimentally accessible probability\textemdash namely,
the statistical probability that if one member of the subensemble
labeled by the pair $a$ and $b$ is randomly selected according to
a uniform probability measure, then the selected member will exhibit
the specific sequence $c_{1},\dots,c_{n}$ of intermediate measurement
outcomes.

As a concrete example, consider an ensemble of $N=10,\!000$ experimental
trials involving a single spin-1/2 degree of freedom that is free
from any nontrivial internal dynamics for the duration of the procedure.
In each trial, the system is prepared in the spin-$z$ eigenstate
$\ket{z+}$, and then is subjected to a spin-$x$ measurement, followed
by a spin-$y$ measurement, and then finally a spin-$z$ measurement.
If one focuses on trials in which the final spin-$z$ measurement
yields $\ket{z-}$, then it follows from the usual algebraic relationships
between the spin-$x$ eigenbasis, the spin-$y$ eigenbasis, and the
spin-$z$ eigenbasis, together with simple arithmetic manipulations,
that the number of trials reduces to approximately $N_{z+,z-}\approx5,\!000$.
It also follows that each of the four possible configurations of intermediate
spin-$x$ and spin-$y$ measurements $\left(x+,y+\right)$, $\left(x+,y-\right)$,
$\left(x-,y+\right)$, and $\left(x-,y-\right)$ shows up in approximately
$1,\!250$ of those $5,\!000$ trials. One can therefore compute the
ratio \eqref{eq:ABLFraction} explicitly for, say, the intermediate
outcome pair $\left(x+,y-\right)$, thereby yielding the answer 
\begin{equation}
\frac{N_{z+,x+,y-,z-}}{N_{z+,z-}}\approx\frac{1,\!250}{5,\!000}=\frac{1}{4}=25\%.\label{eq:ExampleSpinHalfRatio}
\end{equation}

Textbook quantum theory, based on the DvN axioms reviewed in Subsection~\ref{subsec:Textbook-Quantum-Theory},
gives a general theoretical expression for predicting the empirical
probability \eqref{eq:ABLFraction}, called the \textquoteleft ABL
rule.\textquoteright{} Its derivation originally appeared in the ABL
paper, and will be derived in a different way in Subsection~\ref{subsec:A-Derivation-of-the-ABL-Rule}.

\subsection{Boundary Conditions\label{subsec:Boundary-Conditions}}

Newtonian mechanics provides a description of classical objects in
physical space, with trajectories that satisfy dynamical laws consisting
of differential equations that are second-order in time. Because these
dynamical laws are second-order in time, obtaining a definite trajectory
requires specifying both the initial positions of all objects in the
system and also their initial velocities. Once one obtains a definite
trajectory satisfying the dynamical laws, one can ask and answer a
great many questions about the properties of the system at arbitrary
times along that trajectory.

Again because Newton's dynamical laws are second-order in time, one
can instead obtain trajectories by augmenting initial positions with
\emph{final} positions, rather than with initial \emph{velocities},
and then making use of appropriate variational methods. These variational
methods include the principle of least action of Lagrangian mechanics.

In many cases, one obtains a unique trajectory from such a variational
method, when combined with boundary conditions on the past and the
future arrangements of the system. In such a situation, if one knows
both the initial and final arrangements of a system of objects\textemdash that
is, the past and future boundary conditions\textemdash then one is
entitled to make precise inferences about the system at intermediate
times, including inferences about the positions of the system's constituent
objects, their velocities, their kinetic energies, and so forth.

As explained in Subsection~\ref{subsec:A-First-Look-at-the-ABL-Rule},
the experimental protocol for the ABL rule involves a pre-selection
at an initial time as well as a post-selection at a final time. These
two operations might seem analogous to boundary conditions that specify
the initial and final arrangements of objects in Lagrangian mechanics,
but to impose that interpretation would be precisely to engage in
a form of unjustified pattern matching between classical and quantum
concepts\textemdash that is, to impose that interpretation would mean
committing the pattern-matching fallacy \eqref{eq:DefPatternMatchingFallacy}.

It is true that pre-selecting a quantum state is very much like specifying
the initial arrangement of a Newtonian system. However, the rules
for time evolution given by the DvN axioms in Subsection~\ref{subsec:Textbook-Quantum-Theory}
are first-order in time. If one knows the quantum state of a given
system at an initial time, then unitary time evolution as well as
the Born rule and the collapse axioms provide the system's later quantum
states. These time-evolution rules are not second-order in time, and
cannot accommodate the specification of final quantum states as additional
boundary conditions, unless one is willing to violate those time-evolution
rules.

When post-selection is imposed on an ensemble of quantum systems in
the manner in which it is used for the ABL rule, it is due to an external
agent, and represents an abrupt change to the time evolution of each
system that the system has no means of anticipating. The post-selection
is not a boundary condition internal to any particular quantum system
in question, or a boundary condition internal to the ensemble of quantum
systems. In particular, the post-selection is not a boundary condition
in the sense of classical Lagrangian mechanics as merely indicating
where a system's trajectory, on its own, ended up taking the system.
As a consequence, although post-selection might well \emph{reveal}
facts about the past of a particular system or an ensemble of systems,
nothing about each system's quantum state at times before the post-selection
can actually \emph{physically depend} on the later choice of post-selected
quantum state, because any such physical dependence would require
some form of clairvoyance on the part of the system. That is, post-selection
cannot itself be responsible for any past features of the system,
and to assume otherwise would be to commit the post-selection fallacy
\eqref{eq:DefPostSelectionFallacy}.\footnote{In principle, the sort of clairvoyance described here could be permissible
on certain retrocausal interpretations of quantum theory. Alternatively,
on the Everett interpretation (Everett 1956, 1957)\nocite{Everett:1956ttotuwf,Everett:1957rsfqm},
one could argue that post-selection places the observer on a specific
branch of the universal wave function, complete with its own particular
branch-history. Hence, in the case of the Everett interpretation,
the post-selection does not alter the past, but merely singles out
one past from among many others that also exist. Either of these alternative
views, however, would clearly be interpretation-dependent.}

\subsection{A Derivation of the ABL Rule\label{subsec:A-Derivation-of-the-ABL-Rule}}

Deriving the ABL rule is a straightforward exercise in the application
of textbook quantum theory. The derivation presented below also turns
out to be closely related to the derivation of weak values, to be
addressed in other work (Barandes 2026)\nocite{Barandes:2025ttwwv}.

Consider three quantum systems: a subject system to be studied, a
measuring device, and an external agent or observer. The appropriate
Hilbert space for the total system has the tensor-product form 
\begin{equation}
\hilbspace_{\textrm{tot}}=\hilbspace\tensorprod\hilbspace_{\textrm{dev}}\tensorprod\hilbspace_{\textrm{obs}},\label{eq:ABLRuleTotalHilbertSpaceSubjectDeviceObs}
\end{equation}
 where $\hilbspace$, $\hilbspace_{\textrm{dev}}$, and $\hilbspace_{\textrm{obs}}$
are the respective Hilbert spaces for the subject system, the measuring
device, and the observer.

Next, suppose that the initial quantum state $\ket{\Psi_{\textrm{tot}}}$
of the total system assigns a generic state vector $\ket{\Psi}$ to
the subject system, a \textquoteleft ready\textquoteright{} state
vector $\ket{\textrm{dev}\left(\emptyset\right)}$ to the measuring
device, and a \textquoteleft ready\textquoteright{} state vector $\ket{\textrm{obs}\left(\emptyset\right)}$
to the observer: 
\begin{equation}
\ket{\Psi_{\textrm{tot}}}=\ket{\Psi}\tensorprod\ket{\textrm{dev}\left(\emptyset\right)}\tensorprod\ket{\textrm{obs}\left(\emptyset\right)}.\label{eq:ABLRuleTotalStateVectorInitial}
\end{equation}

Let $A,C_{1},\dots,C_{n},B$ be a collection of observables belonging
to the subject system, with respective eigenvalues denoted by the
variables $a,c_{1},\dots,c_{n},b$. Assume that each of these observables
is complete, in the sense that a measurement of any of them fixes
the quantum state of the subject system completely.

Suppose, moreover, that the measuring device and observer are arranged
in advance so that the \emph{observer} measures $A$, then the \emph{measuring
device} measures the chronological sequence $C_{1},\dots C_{n}$,
and then the \emph{observer} measures $B$. Applying unitary time
evolution to the total system, one sees that the first measurement
takes the form 
\begin{equation}
\begin{aligned}\left.\begin{aligned}\ket{\Psi_{\textrm{tot}}} & =\ket{\Psi}\tensorprod\ket{\textrm{dev}\left(\emptyset\right)}\tensorprod\ket{\textrm{obs}\left(\emptyset\right)}\\
 & =\sum_{a}\braket a{\Psi}\ket a\tensorprod\ket{\textrm{dev}\left(\emptyset\right)}\tensorprod\ket{\textrm{obs}\left(\emptyset\right)}\\
\left[\textrm{observer measurement of }A\right] & \mapsto\sum_{a}\braket a{\Psi}\ket a\tensorprod\ket{\textrm{dev}\left(\emptyset\right)}\tensorprod\ket{\textrm{obs}\left(a\right)},
\end{aligned}
\right\} \end{aligned}
\label{eq:ABLRuleFirstMeasurement}
\end{equation}
 where $\textrm{obs}\left(a\right)$ indicates that the observer has
obtained the measurement outcome $a$ for the observable $A$. The
second measurement takes the form 
\begin{equation}
\left.\begin{aligned} & \sum_{a}\braket a{\Psi}\ket a\tensorprod\ket{\textrm{dev}\left(\emptyset\right)}\tensorprod\ket{\textrm{obs}\left(a\right)}\\
 & =\sum_{a}\sum_{c_{1}}\braket a{\Psi}\braket{c_{1}}a\ket{c_{1}}\tensorprod\ket{\textrm{dev}\left(\emptyset\right)}\tensorprod\ket{\textrm{obs}\left(a\right)}\\
\left[\textrm{device measurement of }C_{1}\right] & \mapsto\sum_{a}\sum_{c_{1}}\braket a{\Psi}\braket{c_{1}}a\ket{c_{1}}\tensorprod\ket{\textrm{dev}\left(c_{1}\right)}\tensorprod\ket{\textrm{obs}\left(a\right)},
\end{aligned}
\right\} \label{eq:ABLRuleSecondMeasurement}
\end{equation}
 where $\textrm{dev}\left(c_{1}\right)$ indicates that the measuring
device has obtained the measurement outcome $c_{1}$ for the observable
$C_{1}$. Similarly, after the measuring device carries out its next
measurement, one has 
\begin{equation}
\sum_{a}\sum_{c_{1},c_{2}}\braket a{\Psi}\braket{c_{1}}a\braket{c_{2}}{c_{1}}\ket{c_{2}}\tensorprod\ket{\textrm{dev}\left(c_{1},c_{2}\right)}\tensorprod\ket{\textrm{obs}\left(a\right)},\label{eq:ABLRuleThirdMeasurement}
\end{equation}
 where $\textrm{dev}\left(c_{1},c_{2}\right)$ indicates that the
measuring device has obtained the measurement outcome $c_{1}$ for
the observable $C_{1}$ and then the measurement outcome $c_{2}$
for the observable $C_{2}$, in that chronological order. Continuing,
one ends up finding that the final state vector $\ket{\Psi_{\textrm{tot}}^{\prime}}$
for the total system is given by 
\begin{equation}
\ket{\Psi_{\textrm{tot}}^{\prime}}=\sum_{a,b}\ket b\tensorprod\left[\sum_{c_{1},\dots,c_{n}}\braket b{c_{n}}\braket{c_{n}}{c_{n-1}}\cdots\braket{c_{1}}a\braket a{\Psi}\ket{\textrm{dev}\left(c_{1},\dots,c_{n}\right)}\right]\tensorprod\ket{\textrm{obs}\left(a,b\right)}.\label{eq:TotalStateVectorFinal}
\end{equation}
 Here $\textrm{obs}\left(a,b\right)$ indicates that the observer
has obtained the measurement outcome $a$ for the observable $A$
and $b$ for the observable $B$, whereas $\textrm{dev}\left(c_{1},\dots,c_{n}\right)$
indicates that the measuring device has obtained the chronologically
ordered sequence of measurement outcomes $c_{1},\dots,c_{n}$ for
the respective observables $C_{1},\dots,C_{n}$.

It will be convenient to introduce projection operators 
\begin{equation}
\projector_{a}\defeq\ket a\bra a,\quad\projector_{b}\defeq\ket b\bra b,\quad\projector_{c_{i}}\defeq\ket{c_{i}}\bra{c_{i}}.\label{eq:DefProjectors}
\end{equation}
 Each of these three sets of projection operators gives a complete
and mutually exclusive set, so they each make up a projection-valued-measure
(PVM), in the sense that 
\begin{equation}
\left.\begin{aligned}\projector_{a}\projector_{a^{\prime}}=\delta_{aa^{\prime}}\projector_{a}, & \quad\sum_{a}\projector_{a}=\idmatrix,\\
\projector_{b}\projector_{b^{\prime}}=\delta_{bb^{\prime}}\projector_{b}, & \quad\sum_{b}\projector_{b}=\idmatrix,\\
\projector_{c_{i}}\projector_{c_{i}^{\prime}}=\delta_{c_{i}c_{i}^{\prime}}\projector_{c_{i}}, & \quad\sum_{c_{i}}\projector_{c_{i}}=\idmatrix,
\end{aligned}
\quad\right\} \label{eq:ThreePVMs}
\end{equation}
 where $\idmatrix$ is the identity operator on the subject system's
Hilbert space. (Note that for the third PVM here, the sum is \emph{not}
over the values of $i$ that distinguish the different observables
$C_{1},\dots,C_{n}$, but over the full spectrum $c_{i}$ of eigenvalues
of just the \emph{single} observable $C_{i}$.) One can then write
\eqref{eq:TotalStateVectorFinal} somewhat more compactly as 
\begin{equation}
\ket{\Psi_{\textrm{tot}}^{\prime}}=\sum_{a,b}\sum_{c_{1},\dots,c_{n}}\left[\projector_{b}\projector_{c_{n}}\cdots\projector_{c_{1}}\projector_{a}\ket{\Psi}\right]\tensorprod\ket{\textrm{dev}\left(c_{1},\dots,c_{n}\right)}\tensorprod\ket{\textrm{obs}\left(a,b\right)}.\label{eq:TotalStateVectorFinalCompact}
\end{equation}

Introducing another projection operator 
\begin{equation}
\projector_{\textrm{obs}\left(a,b\right)}\defeq\ket{\textrm{obs}\left(a,b\right)}\bra{\textrm{obs}\left(a,b\right)},\label{eq:DefProjectorObserver}
\end{equation}
 the overall probability $p\left[\textrm{obs}\left(a,b\right)\given\Psi\right]$
for the observer to obtain the specific pair of measurement outcomes
$\left(a,b\right)$, conditioned on the initial state vector $\ket{\Psi}$
of the subject system, then follows from the Born rule: 
\begin{equation}
p\left[\textrm{obs}\left(a,b\right)\given\Psi,\left(A,C_{1},\dots,C_{n},B\right)\right]=\tr\left(\left[\idmatrix\tensorprod\idmatrix_{\textrm{dev}}\tensorprod\projector_{\textrm{obs}\left(a,b\right)}\right]\left[\ket{\Psi_{\textrm{tot}}^{\prime}}\bra{\Psi_{\textrm{tot}}^{\prime}}\right]\right).\label{eq:BornRuleForProbabilityForObserver}
\end{equation}
 Here $\idmatrix_{\textrm{dev}}$ is the identity operator on the
measuring device's Hilbert space. The additional conditioning on $\left(A,C_{1},\dots,C_{n},B\right)$,
as an ordered list, is a reminder that the evolution of the total
system's state vector from $\ket{\Psi_{\textrm{tot}}}$ to $\ket{\Psi_{\textrm{tot}}^{\prime}}$
involved measurements of that specific ordered sequence of observables.
By a straightforward calculation, one finds 
\begin{equation}
p\left[\textrm{obs}\left(a,b\right)\given\Psi,\left(A,C_{1},\dots,C_{n},B\right)\right]=\sum_{c_{1},\dots,c_{n}}\tr\left[\projector_{b}\projector_{c_{n}}\cdots\projector_{c_{1}}\projector_{a}\projector_{\Psi}\projector_{a}\projector_{c_{1}}\cdots\projector_{c_{n}}\right],\label{eq:ProbabilityForObserver}
\end{equation}
 where $\projector_{\Psi}$ is the rank-one density matrix defined
by 
\begin{equation}
\projector_{\Psi}\defeq\ket{\Psi}\bra{\Psi}.\label{eq:DefRankOneDensityMatrix}
\end{equation}

Defining the projection operator 
\begin{equation}
\projector_{\textrm{dev}\left(c_{1},\dots,c_{n}\right)}\defeq\ket{\textrm{dev}\left(c_{1},\dots,c_{n}\right)}\bra{\textrm{dev}\left(c_{1},\dots,c_{n}\right)},\label{eq:DefProjectorDevice}
\end{equation}
the joint probability for the observer to obtain $\left(a,b\right)$
and for the measuring device to obtain the ordered sequence $\left(c_{1},\dots,c_{n}\right)$
likewise follows from the Born rule: 
\begin{equation}
p\left[\textrm{obs}\left(a,b\right),\textrm{dev}\left(c_{1},\dots,c_{n}\right)\given\Psi,\left(A,C_{1},\dots,C_{n},B\right)\right]=\tr\left(\left[\idmatrix\tensorprod\projector_{\textrm{dev}\left(c_{1},\dots,c_{n}\right)}\tensorprod\projector_{\textrm{obs}\left(a,b\right)}\right]\left[\ket{\Psi_{\textrm{tot}}^{\prime}}\bra{\Psi_{\textrm{tot}}^{\prime}}\right]\right).\label{eq:BornRuleForJointProbabilityForObserverAndDevice}
\end{equation}
 The result is 
\begin{equation}
p\left[\textrm{obs}\left(a,b\right),\textrm{dev}\left(c_{1},\dots,c_{n}\right)\given\Psi,\left(A,C_{1},\dots,C_{n},B\right)\right]=\tr\left[\projector_{b}\projector_{c_{n}}\cdots\projector_{c_{1}}\projector_{a}\projector_{\Psi}\projector_{a}\projector_{c_{1}}\cdots\projector_{c_{n}}\right].\label{eq:JointProbabilityForObserverAndDevice}
\end{equation}
 Notice, as expected, that these overall and joint probabilities are
manifestly related by marginalization over the full set of ordered
sequences $\left(c_{1},\dots,c_{n}\right)$: 
\begin{equation}
p\left[\textrm{obs}\left(a,b\right)\given\Psi,\left(A,C_{1},\dots,C_{n},B\right)\right]=\sum_{c_{1},\dots,c_{n}}p\left[\textrm{obs}\left(a,b\right),\textrm{dev}\left(c_{1},\dots,c_{n}\right)\given\Psi,\left(A,C_{1},\dots,C_{n},B\right)\right].\label{eq:ProbabilitiesObserverMarginalizationFromJoint}
\end{equation}

If one applies the DvN collapse axiom to single out the term in the
superposition $\ket{\Psi_{\textrm{tot}}^{\prime}}$ corresponding
to the observer's measurement results $\left(a,b\right)$, then the
reduced density matrix for the measuring device alone is obtained
from the partial trace over the Hilbert spaces of the subject system
and the observer according to 
\begin{equation}
\densitymatrix_{\textrm{dev}\given\textrm{obs}\left(a,b\right),\Psi}=\tr_{\hilbspace,\hilbspace_{\textrm{obs}}}\left(\frac{\left[\idmatrix\tensorprod\idmatrix_{\textrm{dev}}\tensorprod\projector_{\textrm{obs}\left(a,b\right)}\right]\ket{\Psi_{\textrm{tot}}^{\prime}}\bra{\Psi_{\textrm{tot}}^{\prime}}\left[\idmatrix\tensorprod\idmatrix_{\textrm{dev}}\tensorprod\projector_{\textrm{obs}\left(a,b\right)}\right]}{\tr\left(\left[\idmatrix\tensorprod\idmatrix_{\textrm{dev}}\tensorprod\projector_{\textrm{obs}\left(a,b\right)}\right]\ket{\Psi_{\textrm{tot}}^{\prime}}\bra{\Psi_{\textrm{tot}}^{\prime}}\right)}\right).\label{eq:CollapseRuleForReducedDensityMatrixForDevice}
\end{equation}
 The result is a rank-one density matrix that can be expressed in
terms of a state vector given by 
\begin{equation}
\ket{\Psi_{\textrm{dev}\given\textrm{obs}\left(a,b\right),\Psi}}=\frac{\sum_{c_{1},\dots,c_{n}}\bra b\projector_{c_{n}}\cdots\projector_{c_{1}}\projector_{a}\ket{\Psi}\ket{\textrm{dev}\left(c_{1},\dots,c_{n}\right)}}{\sqrt{\sum_{c_{1}^{\prime},\dots,c_{n}^{\prime}}\tr\left(\projector_{b}\projector_{c_{n}^{\prime}}\cdots\projector_{c_{1}^{\prime}}\projector_{a}\projector_{\Psi}\projector_{a}\projector_{c_{1}^{\prime}}\cdots\projector_{c_{n}^{\prime}}\right)}}.\label{eq:CollapsedStateVectorForDevice}
\end{equation}
 If one were to measure the pointer variables of the measuring device
itself, then the Born rule would imply that the corresponding probability
would be 
\begin{equation}
p\left[\textrm{dev}\left(c_{1},\dots c_{n}\right)\given\Psi,\textrm{obs}\left(a,b\right),\left(A,C_{1},\dots,C_{n},B\right)\right]=\frac{\tr\left(\projector_{b}\projector_{c_{n}}\cdots\projector_{c_{1}}\projector_{a}\projector_{\Psi}\projector_{a}\projector_{c_{1}}\cdots\projector_{c_{n}}\right)}{\sum_{c_{1}^{\prime},\dots,c_{n}^{\prime}}\tr\left(\projector_{b}\projector_{c_{n}^{\prime}}\cdots\projector_{c_{1}^{\prime}}\projector_{a}\projector_{\Psi}\projector_{a}\projector_{c_{1}^{\prime}}\cdots\projector_{c_{n}^{\prime}}\right)}.\label{eq:ConditionalProbabilityForDevice}
\end{equation}
 Notice that this conditional probability is related to the joint
probability in \eqref{eq:JointProbabilityForObserverAndDevice} and
the overall probability in \eqref{eq:ProbabilityForObserver} according
to 
\begin{equation}
p\left[\textrm{dev}\left(c_{1},\dots c_{n}\right)\given\Psi,\textrm{obs}\left(a,b\right),\left(A,C_{1},\dots,C_{n},B\right)\right]=\frac{p\left[\textrm{obs}\left(a,b\right),\textrm{dev}\left(c_{1},\dots,c_{n}\right)\given\Psi,\left(A,C_{1},\dots,C_{n},B\right)\right]}{p\left[\textrm{obs}\left(a,b\right)\given\Psi,\left(A,C_{1},\dots,C_{n},B\right)\right]},\label{eq:BayesTheoremForDevice}
\end{equation}
 which is just a form of Bayes' theorem: 
\begin{equation}
p\left(x\given y,z\right)=\frac{p\left(x,y\given z\right)}{p\left(y\given z\right)}.\label{eq:BayesTheoremWithExtraConditional}
\end{equation}

One can simplify the conditional probability \eqref{eq:ConditionalProbabilityForDevice}
by using the identity 
\begin{equation}
\projector_{a}\projector_{\Psi}\projector_{a}=\ket a\braket a{\Psi}\braket{\Psi}a\bra a=\verts{\braket a{\Psi}}^{2}\projector_{a}\label{eq:ProjectorIdentityForSimplifyingABLRule}
\end{equation}
 in both the numerator and the denominator, thereby leading to a cancellation
of a common factor of $\verts{\braket a{\Psi}}^{2}$, assuming that
this factor is nonzero. The conditional probability \eqref{eq:ConditionalProbabilityForDevice}
therefore reduces to the formula 
\begin{equation}
p\left[\textrm{dev}\left(c_{1},\dots c_{n}\right)\given\textrm{obs}\left(a,b\right),\left(A,C_{1},\dots,C_{n},B\right)\right]=\frac{\tr\left(\projector_{b}\projector_{c_{n}}\cdots\projector_{c_{1}}\projector_{a}\projector_{c_{1}}\cdots\projector_{c_{n}}\right)}{\sum_{c_{1}^{\prime},\dots,c_{n}^{\prime}}\tr\left(\projector_{b}\projector_{c_{n}^{\prime}}\cdots\projector_{c_{1}^{\prime}}\projector_{a}\projector_{c_{1}^{\prime}}\cdots\projector_{c_{n}^{\prime}}\right)},\label{eq:CorrectABLRule}
\end{equation}
 which no longer depends on $\ket{\Psi}$. The formula \eqref{eq:CorrectABLRule}
is known as the ABL rule.

\subsection{Time Symmetry\label{subsec:Time-Symmetry}}

Due to the cyclic property of the trace, the ABL rule \eqref{eq:CorrectABLRule}
satisfies the following reverse-ordering symmetry: 
\begin{equation}
p\left[\textrm{dev}\left(c_{1},\dots c_{n}\right)\given\textrm{obs}\left(a,b\right),\left(A,C_{1},\dots,C_{n},B\right)\right]=p\left[\textrm{dev}\left(c_{n},\dots c_{1}\right)\given\textrm{obs}\left(b,a\right),\left(B,C_{n},\dots,C_{1},A\right)\right].\label{eq:ABLRuleReOrderingSymmetry}
\end{equation}
 That is, the conditional probability has the same numerical value
if the measurement sequence $A,C_{1},\dots,C_{n},B$ is carried out
in the opposite chronological order. In the notation of Subsection~\ref{subsec:A-First-Look-at-the-ABL-Rule},
the reverse-ordering symmetry \eqref{eq:ABLRuleReOrderingSymmetry}
of the ABL rule means that the probability $N_{ac_{1}\cdots c_{n}b}/N_{ab}$
appearing in \eqref{eq:ABLFraction} is equal to the probability $N_{bc_{n}\cdots c_{1}a}/N_{ba}$,
with the sequence of the $n+2$ measurements carried out in the opposite
chronological order. 

The ABL paper called this property ``time symmetry.'' Again, the
ABL paper's \emph{title} was ``Time Symmetry in the Quantum Process
of Measurement.''

This notion of time symmetry was incorrect. Carrying out the $n+2$
measurements in the opposite chronological order simply fails to be
the true time-reverse of the experimental protocol. A measurement
process is intrinsically time-directed, as noted by Shimony (1996)\nocite{Shimony:1996abeotsitpom}
and acknowledged by Vaidman (1996b, 1998)\nocite{Vaidman:1996dtsqt,Vaidman:1998tsqt}.
It logically follows that the true time-reverse of the experimental
protocol would entail not only reversing the order of $n+2$ measurements,
but also internally reversing each individual measurement process
itself. 

An analogy might be helpful here. If one buys bread from the store,
and then later one buys fruit from the store, then the time-reverse
of the overall process would not consist of buying fruit first and
then buying bread second. The time-reverse would instead mean something
like \emph{selling }fruit and then later \emph{selling} bread, for
the simple reason that the time-reverse of exchanging money for an
item of food would be exchanging an item of food for money. (Presumably,
the time-reverse would also involve walking backward, speaking backward,
thinking backward, and so forth.) By the same reasoning, the time-reverse
of one measurement followed by a different measurement would not consist
of the \emph{same} pair of measurements merely occurring in the opposite
chronological order, but would instead mean the corresponding \emph{reverse-time
}measurements occurring in the opposite chronological order.

Hence, just as a truly time-symmetric shopping plan would involve
buying food during the first trip and selling food during the second
trip, a truly time-symmetric experimental protocol for a quantum system
would involve carrying out a forward-time measurement at the beginning
and a reverse-time measurement at the end.

After the ABL paper's publication in 1964, subsequent papers continued
to advocate for the ABL paper's incorrect notion of time symmetry.
For example, a 1990 paper by Aharonov and Vaidman made that claim,
even going as far as calling the pre-selection and post-selection
``boundary conditions,'' overlooking the sorts of pattern-matching
problems laid out in Subsection~\ref{subsec:Boundary-Conditions}:
\begin{quotation}
\noindent However, if our task is a description of a quantum system
between two successive measurements, then we know the boundary conditions
in the future as well as in the past. (We assume that both measurements
are complete.) Therefore for the intermediate time interval we have
a complete symmetry under time reversal. The contribution to the description
of the quantum system from the result of the initial measurement is
the usual wave function evolving from the past toward the future,
from the initial measurement to the final measurement. Because of
the symmetry under time reversal, the contribution of the final measurement
should be similar: the wave function evolving backwards in time from
the final measurement to the initial measurement. {[}Aharonov, Vaidman
1990, p. 12{]}\nocite{AharonovVaidman:1990poaqsdttibtm}
\end{quotation}
A 1991 paper by Aharonov and Vaidman contained the following statements
in its concluding section: ``What we have presented here is a novel
approach to standard quantum theory. ... It has an advantage that
it is symmetrical under time reversal.'' (Aharonov, Vaidman 1991,
p. 2327)\nocite{AharonovVaidman:1991cdoaqsaagt}  Papers by other
authors have made similar statements, such as a 2011 paper by Lloyd
et al., which included the following assertion: ``it is a time-symmetrical
formulation of quantum mechanics in which not only the initial state,
but also the final state is specified.'' (Lloyd et al. 2011, p. 025007-5)\nocite{LloydMacconeGarcia-PatronGiovannettiShikano:2011qmotttpst}

\subsection{The ABL Paper\label{sec:The-ABL-Paper}}

The ABL paper's abstract began with a strong claim:
\begin{quotation}
We examine the assertion that the ``reduction of the wave packet,''
implicit in the quantum theory of measurement{[},{]} introduces into
the foundations of quantum physics a time-asymmetric element, which
in turn leads to irreversibility. We argue that this time asymmetry
is actually related to the manner in which statistical ensembles are
constructed. If we construct an ensemble time symmetrically by using
both initial and final states of the system to delimit the sample,
then the resulting probability distribution turns out to be time symmetric
as well. {[}Aharonov, Bergmann, Lebowitz 1964; p. B1410.{]}\nocite{AharonovBergmannLebowitz:1964tsitqpom}
\end{quotation}
On their face, these claims seem doubtful. It is difficult to believe
that one can obtain a time-symmetric formulation of quantum theory
merely by constructing ensembles differently, for reasons already
explained in Subsection~\ref{subsec:Time-Symmetry}. The authors
of the ABL paper did not suggest that they were working outside of
the framework of the DvN axioms, as reviewed in Subsection~\ref{subsec:Textbook-Quantum-Theory},
and for which the reduction or collapse of the quantum state arises
from measurement processes. Again, a generic measurement process is
structured, has a nonzero duration, and, as emphasized by Shimony
(1996)\nocite{Shimony:1996abeotsitpom}, is time-directed. That is,
a measurement process has an intrinsic temporal direction from set-up,
to initiation, to detection, and then to recording. It follows that
whenever one invokes a measurement process, there will be some practical
source of time-asymmetry in the overall system, regardless of how
one tries to set up an ensemble.

Even if one were to attempt to treat measurements as axiomatically
instantaneous events, one would still have to contend with the fact
that a quantum state's time evolution, according to the DvN axioms,
is \emph{discontinuous} to the immediate\emph{ past} of a measurement,
but is \emph{continuous} to the immediate \emph{future} of a measurement,
and leads to measurement-outcome probabilities only in the future
direction. A measurement-induced form of time-asymmetry in the system
is, once again, unavoidable.

Given the thermodynamic-level difficulty of implementing a realistic
measurement process involving macroscopic measuring devices running
in reverse, it is hard to imagine how one could institute a reduction
or collapse of the quantum state at both temporal ends of a duration
of time that could lead to a time-symmetric formulation relevant to
any practical experimental protocol. One would need to construct a
physical ensemble by imposing a forward-time measurement at the beginning
and an infeasible reverse-time measurement at the end, with no clear
way to combine two such opposite-time measurements into a single probability
formula. These basic facts alone present a fundamental obstruction
to the sort of time-symmetric theory that the ABL paper attempted
to formulate. A secondary consequence is that \textquoteleft pre-selection\textquoteright{}
and \textquoteleft post-selection\textquoteright{} are inherently
different from each other in a very physical sense, at a level beyond
merely the fact that one precedes the other in time, as acknowledged
by Vaidman (1996b, 1998)\nocite{Vaidman:1996dtsqt,Vaidman:1998tsqt}.

It follows from this reasoning that the source of the time asymmetry
examined by the ABL paper would not appear to lie in the construction
of ensembles. A corollary is that one should not expect that one could
eliminate that time asymmetry merely by changing how one sets up ensembles.

A couple of paragraphs later in the ABL paper, one finds these statements: 
\begin{quotation}
In this paper we propose to examine the nature of the time symmetry
in the quantum theory of measurement. Rather than delve into the measurement
process itself, which involves a specialized interaction between the
atomic system and a macroscopic device, we shall simply accept the
standard expressions for probabilities of values furnished by the
conventional theory. {[}Aharonov, Bergmann, Lebowitz 1964, p. B1411{]}\nocite{AharonovBergmannLebowitz:1964tsitqpom}
\end{quotation}
It is understandable that one might not wish to get mired in the fine-grained
details of measurement processes, especially given the measurement
problem. However, one must still take into account the fact that a
measurement is not an instantaneous, irreducible, structureless event\textemdash or,
if one were to choose to treat a measurement as if it \emph{were}
instantaneous, then one should be mindful that the measurement separates
discontinuous evolution to the past of a system's quantum state from
continuous evolution to the future of the system's quantum state.
Either way, a measurement is a time-directed process.

The rest of that paragraph went on to say:
\begin{quotation}
\noindent Whereas the conventional theory deals with ensembles of
quantum systems that have been ``preselected'' on the basis of some
initial observation, we shall deduce from it probability expressions
that refer to ensembles that have been selected from combinations
of data favoring neither past nor future. A theory that concerns itself
exclusively with such symmetrically selected ensembles (the ``time-symmetric
theory'') will contain only time-symmetric expressions for the probabilities
of observations. Logically this time-symmetric theory is contained
in the conventional theory but lacks one of the latter's postulates.
{[}Ibid., B1411{]}\nocite{AharonovBergmannLebowitz:1964tsitqpom}
\end{quotation}
Notice that the ABL paper here regarded ensembles selected from ``data
favoring neither past nor future'' as ``symmetrically selected,''
and suggested that an alternative formulation of quantum theory limited
to such ensembles would ``contain only time-symmetric expressions
for the probabilities of observations.'' However, one cannot get
around the need for measurements in pre-selections and post-selections
as long as one is relying on the DvN axioms, or on any other axiomatic
framework for quantum theory that relies on measurements in an essential
way. Again, the ABL paper's interpretation elided the time-directed
nature of the measurements inherent to any such pre-selections or
post-selections. This elision led the ABL paper to employ a notion
of ``time symmetry'' that referred only to changing the sequential
ordering of measurements, rather than truly time-reversing the whole
experimental protocol, including each individual, time-directed measurement
itself. 

The ABL paper assumed that the initial state vector $\ket{\Psi}$
of the subject system was an exact eigenvector $\ket a$ of a given
pre-selected observable. The ABL paper's notation used the labels
$d_{j},\dots,d_{n}$ for the intermediate outcomes instead of $c_{1},\dots,c_{n}$,
used $A$ in place of $\projector_{a}$, used $B$ in place of $\projector_{b}$,
used $D_{i}$ in place of $\projector_{c_{i}}$, and used a diagonal
line $/$ rather than a vertical line $\given$ to denote the \textquoteleft given\textquoteright{}
delimiter. Adjusting the ABL paper's notation to align it with the
notation of the present work, the ABL paper's version of the ABL rule
\eqref{eq:CorrectABLRule} took the form 

\begin{equation}
\left.\begin{aligned}p\left(c_{1},\dots,c_{n}\given a,b\right) & =\frac{p\left(c_{1},\dots,c_{n},b\given a\right)}{p\left(b\given a\right)}\\
 & =\frac{1}{H\left(a,b\right)}\tr\left(\projector_{a}\projector_{c_{1}}\cdots\projector_{c_{n}}\projector_{b}\projector_{c_{n}}\cdots\projector_{c_{1}}\right)\qquad\left[\textrm{ABL's eq. }\left(2.4\right)\right],
\end{aligned}
\quad\right\} \label{eq:ABLRuleDoubleConditionalIncorrect}
\end{equation}
 with $H\left(a,b\right)$ playing the role of $p\left(b\given a\right)$
and determined by overall normalization to be 
\begin{equation}
H\left(a,b\right)=\sum_{c_{1}^{\prime},\dots,c_{n}^{\prime}}\tr\left(\projector_{a}\projector_{c_{1}^{\prime}}\cdots\projector_{c_{n}^{\prime}}\projector_{b}\projector_{c_{n}^{\prime}}\cdots\projector_{c_{1}^{\prime}}\right)\qquad\left[\textrm{ABL's eq. }\left(2.5\right)\right].\label{eq:ABLDoubleConditionalNormalizationIncorrect}
\end{equation}
 Immediately below these formulas, the ABL paper included these statements:
\begin{quotation}
This expression is manifestly time symmetric. If we change the sequence
of measurements to {[}$B,C_{n},\dots,C_{1},A${]}, Eqs. (2.4), (2.5)
remain unchanged. {[}Ibid., p. B1412{]}\nocite{AharonovBergmannLebowitz:1964tsitqpom}
\end{quotation}

The first line of \eqref{eq:ABLRuleDoubleConditionalIncorrect} contained
the ambiguous-looking notation 
\begin{equation}
p\left(c_{1},\dots,c_{n}\given a,b\right)=\frac{p\left(c_{1},\dots,c_{n},b\given a\right)}{p\left(b\given a\right)}.\label{eq:ABLPaperABLRuleProbabilityRatio}
\end{equation}
 This formula made the ABL paper's conditional probability, as it
appears on the left-hand side of \eqref{eq:ABLPaperABLRuleProbabilityRatio},
look like it referred to a logical conjunction of separate propositions
$c_{1},\dots,c_{n}$, to be read as ``$c_{1}$ and $c_{2}$ and $\dots$
and $c_{n}$,'' conditioned on a logical conjunction of separate
propositions $a,b$, to be read as ``$a$ and $b$.'' However, to
assign that meaning or interpretation to the ABL paper's conditional
probability would be to commit the pattern-matching fallacy \eqref{eq:DefPatternMatchingFallacy}.
In particular, one cannot sum on, say, $c_{2}$ to marginalize down
to a shorter measurement sequence that skips the measurement carried
out on the intermediate observable $C_{2}$. The ABL paper's notation
also suppressed the important role played by time-directed measurements
in the overall physical process.

One should compare the ABL paper's formula \eqref{eq:ABLPaperABLRuleProbabilityRatio}
with the more precise (if admittedly more cumbersome) notation \eqref{eq:BayesTheoremForDevice}
from the present work, adapted to the case in which $\ket{\Psi}=\ket a$:
\begin{equation}
p\left[\textrm{dev}\left(c_{1},\dots c_{n}\right)\given a,\textrm{obs}\left(a,b\right),\left(A,C_{1},\dots,C_{n},B\right)\right]=\frac{p\left[\textrm{obs}\left(a,b\right),\textrm{dev}\left(c_{1},\dots,c_{n}\right)\given a,\left(A,C_{1},\dots,C_{n},B\right)\right]}{p\left[\textrm{obs}\left(a,b\right)\given a,\left(A,C_{1},\dots,C_{n},B\right)\right]}.\label{eq:BayesTheoremForDeviceRevisited}
\end{equation}
 This latter version makes clear that $\textrm{dev}\left(c_{1},\dots c_{n}\right)$
is an atomic proposition, and not the logical conjunction of separate
propositions $c_{1},\dots,c_{n}$. Similarly, $\textrm{obs}\left(a,b\right)$
is an atomic proposition, and not the logical conjunction of separate
propositions $a$ and $b$. This latter version also makes manifest
the time-directed nature of the measurements in the experimental protocol.

\subsection{Vaidman's Interpretation\label{subsec:Vaidman's-Interpretation}}

A 1996 paper by Vaidman expressly acknowledged the fundamentally different
roles played by pre-selection and post-selection in the ABL rule:
\begin{quotation}
Note the asymmetry between the {[}pre-selection{]} measurement at
$t_{1}$ and the {[}post-selection{]} measurement at $t_{2}$. Given
an ensemble of quantum systems, it is always possible to prepare all
of them in a particular state $\ket{\Psi_{1}}$, but we cannot ensure
finding the system in a particular state $\ket{\Psi_{2}}$. Indeed,
if the pre-selection measurement yielded a result different from projection
on $\ket{\Psi_{1}}$ we can always change the state to $\ket{\Psi_{1}}$,
but if the measurement at $t_{2}$ did not show $\ket{\Psi_{2}}$,
our only choice is to discard such a system from the ensemble. Note
also the asymmetry of the measurement procedures. The measurement
device has to be prepared before the measurement interaction in the
\textquotedblleft ready\textquotedblright{} state and we cannot ensure
finding the \textquotedblleft ready\textquotedblright{} state after
the interaction. {[}Vaidman 1996b, p. 3{]}\nocite{Vaidman:1996dtsqt}
\end{quotation}
However, shortly thereafter, the paper set aside this concern by asserting
that the only relevant notion of time-symmetry should refer to the
intermediate measurements alone:
\begin{quotation}
\noindent These asymmetries, however, are not relevant to the problem
we consider here. We study the symmetry relative to the measurements
at {[}the intermediate{]} time $t$ for a given pre- and post-selected
system, and we do not investigate the time-symmetry of obtaining such
a system. {[}Ibid., p. 4{]}\nocite{Vaidman:1996dtsqt}
\end{quotation}
The paper restated this assertion, in a section titled ``Time asymmetry
prejudice'': 
\begin{quotation}
In my approach the pre- and post-selected states are given. Only intermediate
measurements are to be discussed. So the frequently posed question
about the probability of the result of the post-selection measurement
is irrelevant. It seems to me that the critics of the time-symmetrized
quantum theory use in their arguments the preconception of an asymmetry.
{[}Ibid., p. 12{]}\nocite{Vaidman:1996dtsqt}
\end{quotation}
These statements represent a departure from the time-symmetric interpretation
obtained from a plain reading of the ABL paper, and an implicit admission
that the time-symmetric interpretation cannot be sustained. 

\subsection{The AAD Paper and the Uncertainty Principle\label{subsec:The-AAD-Paper-and-the-Uncertainty-Principle}}

Recalling the spin-1/2 example presented in Subsection~\ref{subsec:A-First-Look-at-the-ABL-Rule},
consider again an ensemble of $N=10,\!000$ experimental trials involving
a single spin-1/2 degree of freedom, where, in each trial, the system
is prepared in the spin-$z$ eigenstate $\ket{z+}$. Suppose, however,
that the system is then subjected to a spin-$x$ measurement, followed
by a \emph{second} spin-$x$ measurement, and no further measurements.
If one collects only trials in which the final spin-$x$ measurement
yields $\ket{x+}$, then the number of trials reduces to approximately
$N_{z+,x+}\approx5,\!000$. In all these trials, the DvN collapse
axiom trivially implies that the second spin-$x$ measurement has
to give the same result as the first spin-$x$ measurement. Thus,
for this ensemble, the ABL ratio \eqref{eq:CorrectABLRule} trivially
yields 
\begin{equation}
\frac{N_{z+,x+,x+}}{N_{z+,x+}}\approx\frac{5,\!000}{5,\!000}=1=100\%\qquad\left[p\left(x_{+}\given z_{+},x_{+}\right)\textrm{ in the original ABL notation}\right].\label{eq:ModifiedExampleSpinHalfRatio}
\end{equation}
 Notice here the the \emph{middle} measurement value appearing in
this protocol, showing up as the \emph{second} subscript of the numerator
$N_{z+,x+,x+}$, corresponds to a spin-$x$ measurement, in between
the pre-selected spin-$z$ value and the post-selected spin-$x$ value.

If one instead considers an ensemble of $N=10,\!000$ trials in which
the spin-1/2 degree of freedom is prepared in the spin-$z$ eigenstate
$\ket{z+}$, then is subjected to a \emph{second} spin-$z$ measurement,
followed by a spin-$x$ measurement, where one keeps only trials with
the final result $\ket{x+}$, then the number of trials is again approximately
$N_{z+,x+}\approx5,\!000$, for which the DvN collapse axiom ensures
that every second spin-$z$ measurement yields $\ket{z+}$. Hence,
for this ensemble, the ABL ratio \eqref{eq:CorrectABLRule} trivially
gives 
\begin{equation}
\frac{N_{z+,z+,x+}}{N_{z+,x+}}\approx\frac{5,\!000}{5,\!000}=1=100\%\qquad\left[p\left(z_{+}\given z_{+},x_{+}\right)\textrm{ in the original ABL notation}\right].\label{eq:ModifiedExampleSpinHalfRatioAlt}
\end{equation}
 Here the middle measurement value, showing up as the second subscript
of the numerator $N_{z+,z+,x+}$, now corresponds to a spin-$z$ measurement,
in between the pre-selected spin-$z$ value and the post-selected
spin-$x$ value.

Is one justified in concluding from \eqref{eq:ModifiedExampleSpinHalfRatio}
and \eqref{eq:ModifiedExampleSpinHalfRatioAlt} that the spin-$x$
and spin-$z$ observables measured in the middle step of each of these
protocols, despite being represented by noncommuting self-adjoint
operators, both occur with probability $100\%$, in contravention
of the uncertainty principle? The answer would appear to be negative,
because \eqref{eq:ModifiedExampleSpinHalfRatio} and \eqref{eq:ModifiedExampleSpinHalfRatioAlt}
are ensemble properties that refer to fundamentally different ensembles,
so to conclude instead in the affirmative would precisely be to commit
the ensemble fallacy \eqref{eq:DefEnsembleFallacy}.

Nor can one justify claiming that \eqref{eq:ModifiedExampleSpinHalfRatio}
and \eqref{eq:ModifiedExampleSpinHalfRatioAlt} imply a violation
of the uncertainty principle merely because both ratios are amenable
to experimental measurement in the laboratory. To assert the opposite
would be to commit the measurementist fallacy \eqref{eq:DefMeasurementistFallacy}.

Moreover, notice the crucial role played by the post-selection of
$\ket{x+}$. Without that post-selection, both the frequency ratios
\eqref{eq:ModifiedExampleSpinHalfRatio} and \eqref{eq:ModifiedExampleSpinHalfRatioAlt}
would instead have been $5,\!000/10,\!000=50\%$, perfectly in keeping
with the uncertainty principle. The preceding example makes abundantly
clear the perils of trying to make statistical inferences when post-selection
is involved\textemdash and, indeed, when post-selection is invoked
on purpose. This illicit invocation of post-selection is an example
of the post-selection fallacy \eqref{eq:DefPostSelectionFallacy},
and gives yet another reason to doubt that the preceding example represents
a true exception to the uncertainty principle.

However, a 1985 paper by Albert, Aharonov, and D'Amato (AAD), titled
``Curious New Statistical Prediction of Quantum Mechanics'' and
appearing in the journal \emph{Physical Review Letters}, (Albert,
Aharonov, D'Amato 1985)\nocite{AlbertAharonovDAmato:1985cnspoqm},
argued that the ABL rule provided a way to violate the uncertainty
principle in just this manner:\footnote{The quoted passage attributes these claims to the authors of the original
ABL paper. However, As Sharp and Shanks later pointed out, there does
not appear to be evidence that these views were expressed in the original
1964 ABL paper: ``Actually, Albert et al. attribute these conclusions
to Aharonov et al. (1964) but we can find no evidence of either conclusion
in that work.'' (Sharp, Shanks 1993, p. 494, footnote 2)\nocite{SharpShanks:1993trafotsqm}
The AAD paper may have been referring to a different paper, published
the previous year (Aharonov, Albert 1984)\nocite{AharonovAlbert:1984itunoteafqmsi}.} 
\begin{quotation}
\noindent Suppose that {[}a given quantum{]} system is measured at
time $t_{i}$ to be in the state $\ket{A=a}$ (where $A$ represents
some complete set of commuting observables of the system, and $a$
represents some particular set of eigenvalues of those observables),
and is measured at time $t_{f}$ ($t_{f}>t_{i}$) to be in the state
$\ket{B=b}$. What do these results imply about the results of other
experiments that might have been carried out within the interval $\left(t_{i}<t<t_{f}\right)$
between them? It turns out that the probability (which was first written
down by Aharonov, Bergmann, and Lebowitz) that a measurement of some
complete set of observables $C$ within that interval, \emph{if} it
were carried out, would find that $C=c_{j}$ is 
\begin{equation}
P\left(c_{j}\right)=\frac{\verts{\braket{A=a}{C=c_{j}}}^{2}\verts{\braket{C=c_{j}}{B=b}}^{2}}{\sum_{i}\verts{\braket{A=a}{C=c_{i}}}^{2}\verts{\braket{C=c_{i}}{B=b}}^{2}}\qquad\left[\textrm{AAD's eq. }\left(1\right)\right];\label{eq:AADVersionOfABLRule}
\end{equation}
 and that formula entails, among other things, that $P\left(a\right)=P\left(b\right)=1$.
Consequently, these authors maintain that such a system, within such
an interval, must have definite, dispersion-free values of both $A$
and $B$, whether or not $A$ and $B$ may happen to commute. In their
view, the proper quantum mechanical descriptions of the past and the
future are essentially different: Our knowledge of the past is not
restricted, in the same way as our ability to predict the future,
by the uncertainty relations; indeed, so far as the past is concerned,
the quantal formalism itself \emph{requires} that those relations
be violated. {[}Ibid., pp. 5\textendash 6, emphasis in the original{]}\nocite{AlbertAharonovDAmato:1985cnspoqm}
\end{quotation}
The AAD paper acknowledged that claims of dispersion-free values of
non-commuting observables, and violations of the uncertainty principle,
might sound surprising, in light of various no-go theorems. The AAD
paper nevertheless doubled down:
\begin{quotation}
Is it somehow mistaken, then, or somehow misleading, to suppose that
(1) attributes definite values to $A$ and $B$? Is it that (1) itself
produces some contradiction? How? Where?

No. It turns out (and this is the subject of the present note) that
there is a remarkable and heretofore unknown property of the quantal
statistics whereby quantum mechanical systems, within the interval
between two measurements, \emph{fail} to satisfy that assumption (the
assumption about the projection operators), and so evade its consequences.
{[}Ibid., p. 6, emphasis in the original{]}\nocite{AlbertAharonovDAmato:1985cnspoqm}
\end{quotation}

Once again, this erroneous conclusion may have been precipitated by
an overly minimalist choice of notation. In the more expansive and
precise notation of \eqref{eq:BayesTheoremForDevice}, the probability
\eqref{eq:AADVersionOfABLRule} in the AAD paper would have been written
instead as 
\begin{equation}
p\left[\textrm{dev}\left(c_{j}\right)\given a,\textrm{obs}\left(a,b\right),\left(A,C,B\right)\right]=\frac{p\left[\textrm{obs}\left(a,b\right),\textrm{dev}\left(c_{j}\right)\given a,\left(A,C,B\right)\right]}{p\left[\textrm{obs}\left(a,b\right)\given a,\left(A,C,B\right)\right]}.\label{eq:CorrectAADProbability}
\end{equation}

The operational meaning of the probability \eqref{eq:CorrectAADProbability},
as outlined in Subsection~\ref{subsec:A-First-Look-at-the-ABL-Rule},
concerns the fraction of the appropriately constructed ensemble whose
members show the intermediate measurement result $c_{j}$ for the
complete set of observables represented by $C$. The choice of $C$
for the intermediate measurement is central to defining the physical
ensemble in question. 

To ask instead for the probability that an intermediate measurement
of $A$ should yield $a$ would mean to construct a \emph{different}
experimental protocol producing a \emph{different} physical ensemble,
for which the ratio yields $1$: 
\begin{equation}
p\left[\textrm{dev}\left(a\right)\given a,\textrm{obs}\left(a,b\right),\left(A,A,B\right)\right]=\frac{p\left[\textrm{obs}\left(a,b\right),\textrm{dev}\left(a\right)\given a,\left(A,A,B\right)\right]}{p\left[\textrm{obs}\left(a,b\right)\given a,\left(A,A,B\right)\right]}=1.\label{eq:CorrectAADTrivialProbabilityFirst}
\end{equation}
 Similarly, to ask for the probability that an intermediate measurement
of $B$ should yield $b$ would mean to construct yet \emph{another}
experimental protocol producing a distinct physical ensemble of its
own, for which the ratio again yields $1$: 
\begin{equation}
p\left[\textrm{dev}\left(b\right)\given a,\textrm{obs}\left(a,b\right),\left(A,B,B\right)\right]=\frac{p\left[\textrm{obs}\left(a,b\right),\textrm{dev}\left(b\right)\given a,\left(A,B,B\right)\right]}{p\left[\textrm{obs}\left(a,b\right)\given a,\left(A,B,B\right)\right]}=1.\label{eq:CorrectAADTrivialProbabilitySecond}
\end{equation}
 The three probabilities \eqref{eq:CorrectAADProbability}, \eqref{eq:CorrectAADTrivialProbabilityFirst},
and \eqref{eq:CorrectAADTrivialProbabilitySecond} here refer to three
separate physical ensembles, and so it would be a mistake to try to
draw inferential conclusions about any one of them from either or
both of the other two. As explained earlier, it is also dangerous
to jump to conclusions about statistical inferences when one's set-up
involves post-selection, on pain of committing the post-selection
fallacy \eqref{eq:DefPostSelectionFallacy}. 

To make completely clear why the AAD authors did not find a loophole
in the uncertainty principle, it will be useful to present an operational
argument for how one can experimentally verify the uncertainty principle.
One imagines setting up \emph{one} physical ensemble of identical
quantum systems whose definition consists solely of a preparation
or \textquoteleft pre-selection\textquoteright{} of each member of
the ensemble in the same initial quantum state $\densitymatrix$.
One then carries out a measurement of some observable $C$ for half
of the members of the ensemble, and a measurement of some other observable
$D$ for the other half of the members of the ensemble, where $C$
and $D$, as self-adjoint operators on the system's Hilbert space,
may fail to commute, in the sense that $CD-DC\ne0$. That is, one
carries out a \emph{controlled experiment} by \emph{fixing} $\densitymatrix$
as the definition of the entire physical ensemble and then \emph{independently}
\emph{varying} just the choice of $C$ or $D$ within that fixed ensemble,
\emph{without} any post-selection. One finds that the respective spreads
or standard deviations $\Delta C$ and $\Delta D$ of measurement-outcome
distributions for $C$ and $D$ have a product $\Delta C\,\Delta D$
that satisfies the inequality 
\begin{equation}
\Delta C\,\Delta D\geq\frac{1}{2}\verts{\tr\left[\left(CD-DC\right)\densitymatrix\right]},\label{eq:UncertaintyPrinciple}
\end{equation}
 which is just the uncertainty principle. In some cases, such as for
a particle's position $C=x$ and corresponding momentum $D=p_{x}$,
the right-hand side of \eqref{eq:UncertaintyPrinciple} reduces to
$\hbar/2$, and it is impossible for $C$ and $D$ both to have vanishing
dispersion, regardless of the initial quantum state $\densitymatrix$.

For the AAD argument, by contrast, one not only explicitly carries
out a form of post-selection, but one must also have \emph{two} physical
ensembles produced according to \emph{different} experimental protocols:
a first ensemble produced by measuring the chronological sequence
$A,C,B$ and keeping only members of the ensemble for which $A=a$
and $B=b$, and a second ensemble produced by measuring the chronological
sequence $A,D,B$ and similarly keeping only members for which $A=a$
and $B=b$. Crucially, notice that one cannot include $B$ in the
experimental protocols that produce these two ensembles without also
including either $C$ or $D$, in contrast with the single ensemble
constructed for the uncertainty principle in the previous paragraph.
For each of these two ensembles, the physical post-selection process
for $B$ implicitly depends on the previous measurement of $C$ (for
the first ensemble) or on the previous measurement of $D$ (for the
second ensemble). Thus, one cannot imagine replacing $C$ with $D$
without \emph{also} changing the physical post-selection process for
$B$ as well. It is true that replacing $C$ with $D$ does not alter
the pre-selection on $A$, but to make the same assumption about the
post-selection on $B$ would be precisely to assume the erroneous
form of time-symmetry for quantum measurements that the present paper
rigorously argued against in Subsection~\ref{subsec:Time-Symmetry}.

As a consequence, for the AAD argument, it is not possible to carry
out a controlled experiment by \emph{fixing} the pre-selection on
$A$ and the post-selection on $B$ to produce a single physical ensemble,
and then \emph{independently} \emph{varying} the choice of $C$ or
$D$ to obtain a violation of the uncertainty principle. This lack
of independence between the choice of intermediate observable $C$
or $D$ and the post-selection on $B$ is concealed by the notation
$P\left(c_{j}\right)$ used in the AAD version of the ABL rule, \eqref{eq:AADVersionOfABLRule},
as well as in the notation $p\left(c_{1},\dots,c_{n}\given a,b\right)$
used in the original ABL rule \eqref{eq:ABLRuleDoubleConditionalIncorrect},
but is manifest in the more precise notation used in this paper's
formula for the ABL rule in \eqref{eq:CorrectABLRule}.

The present discussion highlights another important feature of textbook
quantum theory that is clear from the review of the DvN axioms in
Subsection~\ref{subsec:Textbook-Quantum-Theory}: at least in some
cases, the textbook theory can make concrete, reliable predictions
for a \emph{single} quantum system, without the need to invoke an
ensemble. This feature is due, in part, to the fact that the DvN axioms
prescribe that observables are assigned at the single-system level,
not at an ensemble level. It therefore makes sense to talk about measuring
observables for a single system, to regard measurement outcomes as
statements about that single system, and to regard measurement probabilities
and expectation values as \emph{referring} to observables of that
single system. (This key feature of textbook quantum theory also turns
out to have important implications for weak values, to be discussed
in other work.)

As an example that is highly relevant to the uncertainty principle,
consider a single system with two noncommuting observables $C$ and
$D$. If one repeatedly measures $C$, without giving the system significant
time for internal evolution between the measurements, then the DvN
collapse axiom will ensure that one reliably obtains the same result
each time, within reasonably small error bars. However, an intervening
measurement of the observable $D$ will generically lead to a significant
change in subsequent measurements of $C$, to a degree controlled
in part by the commutator $CD-DC$. This \emph{experimental} noncommutativity
between \emph{algebraically} noncommuting observables $C$ and $D$
is perhaps the most universal feature of all quantum systems, and
provides a concrete, operational meaning to the uncertainty principle
at the single-system level. In particular, a core part of the uncertainty
principle is precisely that it has this implication for a single system.

The AAD argument, by contrast, always requires (multiple) ensembles,
so it refers exclusively to patterns of behavior in ensembles. As
a consequence, the AAD argument has no meaning \emph{in principle}
for a single system, and therefore inevitably misses a core part of
the uncertainty principle. Thus, \emph{a fortiori}, one cannot interpret
the AAD argument as giving any real insight into the uncertainty principle
for a single system, let alone providing a loophole, without committing
the ensemble fallacy \eqref{eq:DefEnsembleFallacy}.

Over the years since the AAD paper was published, several other papers
have pointed out, in more narrow ways, the failure of the AAD paper
to show that the ABL rule supports counterfactual reasoning. For instance,
in a 1986 paper, Bub and Brown wrote:
\begin{quotation}
That {[}the AAD{]} argument is fallacious can be seen by noting that
the subensemble of the preselected {[}$a${]} ensemble that is post-selected
for {[}$b${]} via an intervening {[}$C${]} measurement differs from
the subensemble that is post-selected for {[}$b${]} via an intervening
{[}$D${]} measurement. {[}Bub, Brown 1986, p. 2338{]}\nocite{BubBrown:1986cpoqewhbbpaps}
\end{quotation}
Sharp and Shanks made a similar case in a 1993 paper, and came to
a conclusion not very different from the one reached in the present
work: 
\begin{quotation}
Interpreted correctly, the ABL-Rule makes predictions only as to the
outcomes of actual measurements conducted upon systems subject to
both and pre- and post-selection. So interpreted, the rule is not
in conflict with orthodox quantum mechanics, but neither does it yield
fresh insights about the fundamental interpretive issues in quantum
mechanics. {[}Sharp, Shanks 1993, p. 499{]}\nocite{SharpShanks:1993trafotsqm}
\end{quotation}
One finds similar criticisms over the use of the ABL rule for counterfactuals
in papers concerned with the consistent-histories interpretation of
quantum theory (Griffiths 1984, Cohen 1995, Kastner 1999)\nocite{Griffiths:1984chatioqm,Cohen:1995papqscmach,Kastner:1999ttbpaortrtcuotar}.

\section{Conclusion\label{sec:Conclusion}}

The present work has reviewed the ABL rule, and raised several challenges
to some of the ways that it has been interpreted in the research literature.
In particular, this paper has argued that the ABL rule does not provide
a time-symmetric formulation of quantum theory and does not lead to
true violations of the uncertainty principle. Along the way, this
paper has highlighted several relevant fallacies that are relevant
to this critical analysis, including the ensemble fallacy \eqref{eq:DefEnsembleFallacy},
the post-selection fallacy \eqref{eq:DefPostSelectionFallacy}, the
pattern-matching fallacy \eqref{eq:DefPatternMatchingFallacy}, and
the measurementist fallacy \eqref{eq:DefMeasurementistFallacy}.

Are these challenges worth discussing? Does it really matter if the
ABL rule is not really time-symmetric, or if post-selection can lead
to erroneous statistical inferences? The answer to these questions
is yes, not only for reasons of philosophical rigor, but also because
physics research since 2000 has made substantial use of these interpretational
claims about the ABL rule and post-selection, and that research merits
scrutiny. In particular, other work will investigate the connections
between the ABL rule and weak values (Barandes 2026)\nocite{Barandes:2025ttwwv}.

At a broader level, this paper should be taken as an argument against
the cavalier use of post-selection to generate publication-worthy
results, and as a call for research journals to insist on more careful
explanations by authors that when they use post-selection, their results
are actually produced by quantum theory itself, and are not merely
manifestations of selection bias. It is heartening that this important
form of self-corrective introspection is already happening in the
research literature (see, for example, Wharton, Price 2025)\nocite{WhartonPrice:2025bivweisjp}.

\section*{Acknowledgments}

The author would especially like to thank David Albert, David Kagan,
Logan McCarty, Xiao-Li Meng, John Norton, James Robins, Yossi Sirote,
Kelly Werker Smith, Lev Vaidman, Larry Wasserman, and Ken Wharton
for helpful discussions.

\bibliographystyle{1_home_jacob_Documents_Work_My_Papers_2025-The_ABL_Rule_and_Weak_Values_custom-abbrvalphaurl}
\bibliography{0_home_jacob_Documents_Work_My_Papers_Bibliography_Global-Bibliography}

\end{document}